\documentclass[sort&compress]  {aipproc}
\layoutstyle{6x9}

\newcommand{\apj}{Astrophys. J.}
\newcommand{\mnras}{Mon. Not. R. Astron. Soc.}

\newcommand{\prd}{Phys. Rev. D}
\newcommand{\prl}{Phys. Rev. Lett}
\newcommand{\lsim}{\mathrel{\hbox{\rlap{\lower.55ex\hbox{$\sim$}} \kern-.3em \raise.4ex \hbox{$<$}}}}
\newcommand{\gsim}{\mathrel{\hbox{\rlap{\lower.55ex\hbox{$\sim$}} \kern-.3em \raise.4ex \hbox{$>$}}}}

\begin{document}

\title{"Thermal axion constraints in non-standard thermal histories}


\author{Daniel Grin}{
  address={California Institute of Technology, Mail Code 130-33, Pasadena, CA 91125}
}
\author{Tristan Smith}{
  address={California Institute of Technology, Mail Code 130-33, Pasadena, CA 91125}
}
\author{Marc Kamionkowski}{
  address={California Institute of Technology, Mail Code 130-33, Pasadena, CA 91125}
}

\begin{abstract}
 There is no direct evidence for radiation domination prior to big-bang nucleosynthesis, and so it is useful to consider how constraints to thermally-produced axions change in non-standard thermal histories. In the low-temperature-reheating scenario, radiation domination begins as late as $\sim 1$ MeV, and is preceded by significant entropy generation. Axion abundances are then suppressed, and cosmological limits to axions are significantly loosened. In a kination scenario, a more modest change to axion constraints occurs. Future possible constraints to axions and low-temperature reheating are discussed.
\end{abstract}

\maketitle

\section{Introduction}
If the axion has mass $m_{\rm a}\gsim 10^{-2}~{\rm eV}$, it will be produced thermally, with cosmological abundance $\Omega_{{\rm a}}h^{2}=\left(m_{{\rm a}}/130~{\rm eV}\right)\left(10/g_{*\rm F}\right)\label{thermalfreeze},$ where $g_{*{\rm F}}$ is the effective number of relativistic degrees of freedom when axions freeze out \cite{crewther, baluni,pq,changchoi,notinvisible,kt,murayama}. If $m_{\rm a}\lsim 1~{\rm eV}$, axions free-stream to erase density perturbations while they are relativistic, and thus suppress the matter power spectrum on small scales, much like neutrinos \cite{hu1,hu2,kt,haneutrino5,changchoi,crotty,tegmarksdss,barger,seljakneutrino,fukugitawmap3,spergelwmap3,pierpaoli}. Data from large-scale structure (LSS) surveys and cosmic microwave-background (CMB) observations impose the constraint $m_{\rm a}\lsim1~{\rm eV}$ to light hadronic axions \cite{hanraffelt,raffeltwong,slosar}. We restrict our attention to hadronic axions in this work.

These constraints were determined in the standard radiation-dominated scenario. The transition to radiation domination after inflation might be gradual \cite{kamion}. In a modified thermal history, relic abundances may change, due to modified freeze-out temperatures and suppression from entropy generation.

The universe could have reheated to a temperature as low as $1~{\rm MeV}$ \cite{kawasaki2,ichikawa3,giudice2,hannestadreheat,kolbnotario}. This low-temperature reheating (LTR) scenario may be modeled simply through the decay of a massive particle $\phi$ into radiation, with fixed rate $\Gamma_{\phi}$. This decay softens the scaling of temperature $T$ with cosmological scale factor $a$, increasing the Hubble parameter $H(T)$ and leading to earlier freeze-out for certain relics. Entropy generation then highly suppresses these relic abundances. Kination models offer another alternative to the standard thermal history, without entropy production, and cause more modest changes in abundances \cite{kination}.

Past work has determined the relaxation in constraints to neutrinos, weakly interacting massive particles, and non-thermally produced axions are relaxed in LTR \cite{giudice1,giudice2,yaguna}. Here, we 
present new constraints to thermally-produced axions in the LTR scenario. We point the reader to Ref. \cite{reh} for a discussion of the more modest changes to axion constraints in the kination scenario, and for additional details relevant to the following discussion. We conclude by discussing the impact of future LSS surveys and CMB measurements of the primordial helium abundance on the allowed parameter space for axions.

\section{Low-temperature reheating (LTR)}
\label{ltrsec}
In the LTR scenario, the density of $\phi$ particles and radiation obey \cite{chung,giudice1,giudice2}:
\begin{equation}
\frac{1}{a^{3}}\frac{d\left(\rho_{\phi}a^{3}\right)}{dt}=-\Gamma_{\phi}\rho_{\phi}~~~~
\frac{1}{a^{4}}\frac{d\left(\rho_{R}a^{4}\right)}{dt}=\Gamma_{\phi}\rho_{\phi},\label{axabev}\end{equation} where $\rho_{\phi}$ and $\rho_{R}$ denote the energy densities in the scalar field and radiation, and $a$ is the cosmological scale factor, whose evolution is given by the Friedmann equation. The reheating temperature $T_{\rm rh}$ is defined by $\Gamma_{\phi}\equiv \sqrt{4\pi^{3}g_{*\rm rh}/45}~T_{{\rm rh}}^{2}/M_{{\rm pl}}$ \cite{giudice1,chung,kt}, where $M_{{\rm pl}}$ is the Planck mass and $g_{*\rm rh}$ is the value of $g_{*}$ when $T=T_{\rm rh}$. 

At the beginning of reheating, $\phi$ dominates the energy density. The temperature is related to the radiation energy density by $T\propto\rho_{R}^{1/4}$  \cite{kt} . We integrate Eqs.~(\ref{axabev}) to obtain the dependence of $T$ on $a$ \cite{reh}. When the scalar begins to decay, the temperature rises quickly to a maximum and then falls as $T\propto a^{-3/8}$. This shallow scaling of temperature with scale factor results from the transfer of scalar-field energy into radiation. When $\rho_{\rm R}$ overtakes $\rho_{\phi}$ near $T\sim T_{\rm rh}$, the epoch of radiation domination begins, with the usual $T\propto a^{-1}$ scaling \cite{reh}. 

During reheating,  $H(T)\propto \left(T/T_{\rm rh}\right)^{2}T^{2}/M_{\rm pl}$ \cite{giudice1,giudice2}, the universe thus expands faster than during radiation domination, and the equilibrium condition $\Gamma\equiv n\left\langle\sigma v\right\rangle\gsim H$ is harder to meet. Relics with freeze-out temperature $T_{\rm F}\geq T_{\rm max}$ will thus have suppressed abundances because they never come into chemical equilibrium. Relics with $T_{\rm rh}\lsim T_{\rm F}\lsim T_{\rm max}$ come into chemical equilibrium, but their abundances are reduced by entropy production. 

\section{Axion production}
\label{production}
\begin{figure}[ht]
\includegraphics[width=5.8in]{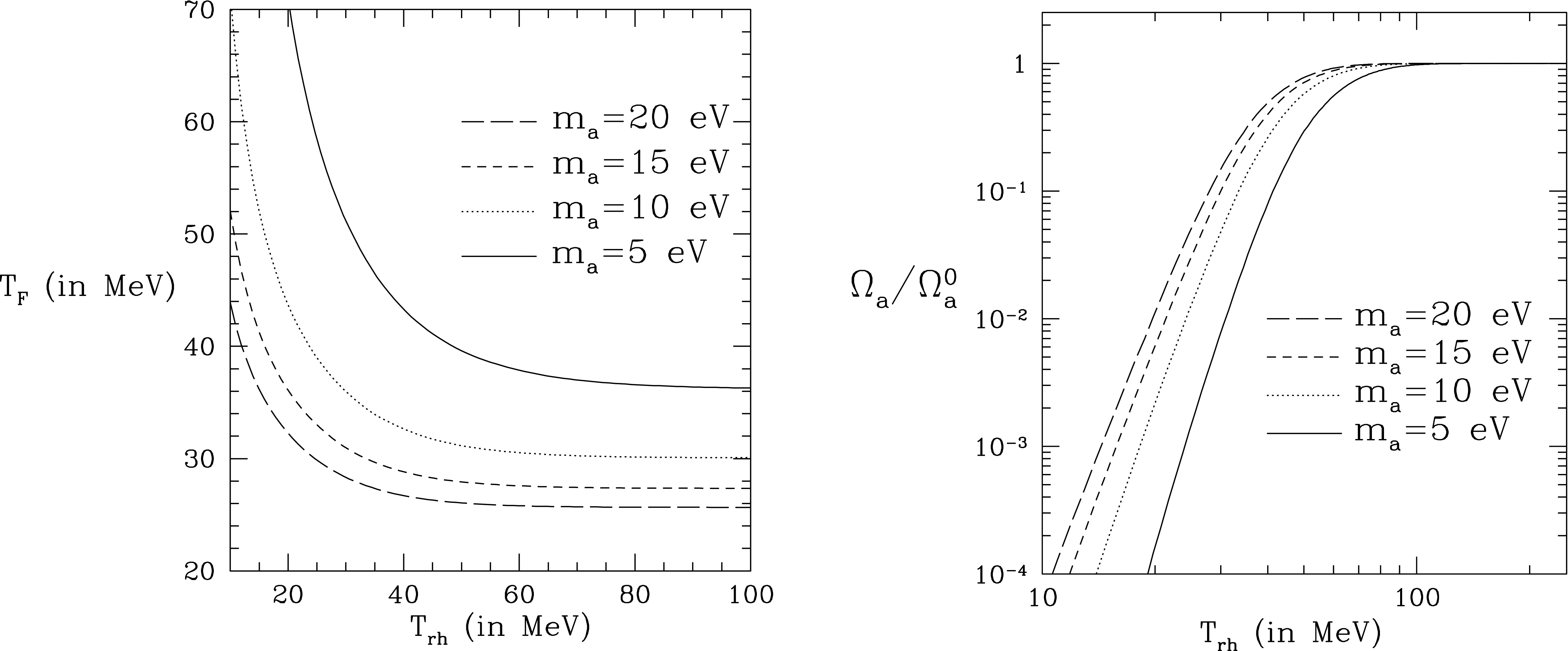}
\caption{The left panel shows the freeze-out temperature of the reactions $\pi^{+}+\pi^{-}\leftrightarrow \pi^{0}+a$, $\pi^{+}+\pi^{0}\leftrightarrow \pi^{+}+ a$, and $\pi^{-}+{\pi^{0}}\leftrightarrow \pi^{-}+a$, as a function of $T_{\rm rh}$, for $4$ different axion masses. The right panel shows the axion abundance $\Omega_{\rm a}$ normalized by its standard value $\Omega_{\rm a}^{0}$ for $4$ different axion masses.}
\label{ft_dependence}
\end{figure}

Standard hadronic axions with $m_{{\rm a}}\gsim 10^{-2}~{\rm eV}$ are produced by the channels $\pi^{+}+\pi^{-}\to  a+\pi^{0}$, $\pi^{+}+\pi^{0}\to \pi^{+}+ a$, and $\pi^{-}+\pi^{0}\to a+\pi^{-}$\cite{hanraffelt,changchoi,kt,khlopovonprod}. Numerically evaluating the expression from Ref. \cite{changchoi} for the axion-production rate $\Gamma$ and solving Eq.~(\ref{axabev}) for $H\left( T\right)$, we estimate the axion freeze-out temperature $T_{\rm F}$ using the condition $\Gamma\left(T_{\rm F}\right)\sim H\left(T_{\rm F}\right)$. As $T_{\rm rh}$ is lowered, axions freeze out earlier due to the higher value of $H$, as shown in Fig.~\ref{ft_dependence}. As $T_{\rm rh}$ increases, the $T\propto a^{-3/8}$ epoch becomes less relevant, and $T_{\rm F}$ asymptotes to its standard value.  Now, since $\Gamma \propto f_{\rm a}^{-2}\propto m_{\rm a}^{2}$ \cite{changchoi}, higher-mass axions keep up with the Hubble expansion for longer and generally decouple at lower temperatures. 

The resulting axion abundance is \cite{reh}
\begin{equation}
\Omega_{\rm a}h^{2}=\frac{m_{\rm a, \rm eV}}{130}\left(\frac{10}{g_{*\rm F}}\right)\gamma\left(T_{\rm rh}/T_{\rm F}\right)~~~~~\gamma(\beta)\sim
\left\{\begin{array}{ll}
\beta^{5}\left(\frac{g_{* \rm rh}}{g_{*\rm F}}\right)^{2}\left(\frac{g_{*\rm F}}{g_{*\rm rh}}\right)&\mbox{if $\beta\ll1$,}\\ 
1&\mbox{if $\beta\gg 1$,}
\end{array}\label{newomew}\right.
\end{equation} where $m_{\rm a,\rm eV}$ is the axion mass in units of ${\rm eV}$. 

When $T_{\rm rh}\lsim T_{\rm F}$, the present mass density in axions is severely suppressed, because of entropy generation. Using the numerical solution for $a\left(T\right)$, we obtain $\Omega_{\rm a}$. In the right panel of Fig.~\ref{ft_dependence},  we show $\Omega_{\rm a}$ normalized by its standard value, $\Omega_{\rm a}^{0}$, as a function of $T_{\rm rh}$. For $T_{\rm rh}\gg T_{\rm F}$, the axion abundance asymptotes to $\Omega_{\rm a}^{0}$.  

\section{Constraints to axions}
Most constraints to the axion mass come from its two-photon coupling $g_{a \gamma \gamma}$ \cite{Raffelt,kt,kaplan,srednicki,kephartw,Bershady:1990sw,notinvisible,ressellt,grin,gnedin,cast,sikivie_cavity}. This coupling depends on the up-down quark mass ratio $r$, for which there are experimentally allowed such that $g_{a \gamma \gamma}$ vanishes, and so constraints to axions from star clusters, helioscope, RF cavity, and telescope searches may all be lifted \cite{buckleymurayama,murayama}. In contrast, the hadronic couplings do not vanish for any experimentally allowed $r$ values. Axion searches based on these couplings are underway, and have already imposed the $m_{\rm a}\lsim~1~{\rm keV}$ range \cite{kekez}.  These couplings also determine the relic abundance of axions, and so constraints may be obtained from cosmology.  

\begin{figure}[t]
\includegraphics[width=5.65in]{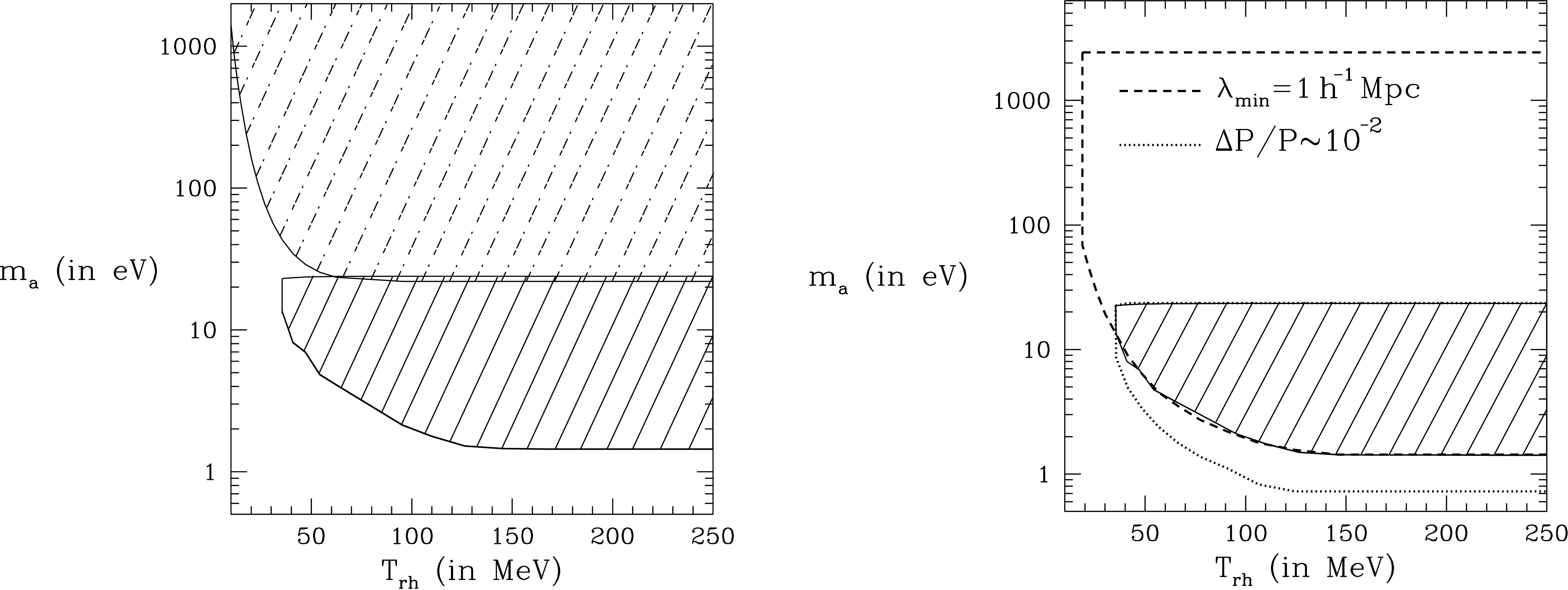}
\caption{The left panel shows upper limits to $m_{\rm a}$ in the LTR scenario. The dot-dashed hatched region shows the region excluded by the constraint $\Omega_{{\rm a}}h^{2}<0.135$. The solid hatched region shows the axion parameter space excluded by WMAP1/SDSS data. At low $T_{\rm rh}$, limits to the axion mass are loosened.  The right panel shows the estimated sensitivity  possible with LSST measurements of the matter power spectrum \cite{lsstintro,lsstzhan}, or from Lyman-$\alpha$ forest measurements of clustering on smaller length scales \cite{viel}. 
The hatched region indicates the region excluded by WMAP1/SDSS measurements.}
\label{limits}
\end{figure}

Mass constraints to thermal axions from cosmology are considerably relaxed because of entropy generation. A conservative constraint is obtained by requiring that axions not exceed the matter density of $\Omega_{\rm m}h^{2}\simeq 0.135$ \cite{spergelwmap3} and is shown by the dot-dashed hashed region in Fig.~\ref{limits}.  If $T_{\rm rh}\lsim 40~{\rm MeV}$, constraints are considerably relaxed.  When $T_{\rm rh}\gsim 95~{\rm MeV}$, we obtain $m_{\rm a}\lsim 22~{\rm eV}$, equal to the standard result. 

Axions will free stream at early times, decreasing the matter power spectrum on length scales smaller than the comoving free-streaming scale, evaluated at matter-radiation equality:
\begin{eqnarray}\lambda_{\rm fs}\simeq \left(196~{\rm Mpc}/m_{\rm a, \rm eV}\right)\left(T_{\rm a}/T_{\nu}\right) \left\{1+\ln{\left[0.45 m_{\rm a, \rm eV}\left(T_{\nu}/T_{\rm a}\right)\right]}\right\}.\label{reallfs}\end{eqnarray}

This suppression is given by $\Delta P/P\simeq-8\Omega_{\rm a}/\Omega_{m}$ if $\Omega_{\rm a}\ll \Omega_{m}$ \cite{hanraffelt,raffeltwong,hann5} and imposes a constraint to $\Omega_{\rm a}h^{2}$. Including entropy generation, the relationship between the effective axion temperature $T_{\rm a}$ and the neutrino temperature $T_{\rm \nu}$ is
\begin{equation}
\frac{T_{a}}{T_{\nu}}\simeq \left[\frac{11}{4}\left(\frac{T_{\rm rh}}{T_{\rm F}}\right)^{5}\left(\frac{g_{*\rm rh}g_{*0}}{g_{*\rm F}^{2}}\right)\right]^{1/3}~\mbox{if $T_{\rm F}\ge T_{\rm rh}$,}~
\frac{T_{\rm a}}{T_{\rm \nu}}\simeq(10.75/g_{*\rm F})^{1/3}~\mbox{if $T_{\rm F}<T_{\rm rh}$.}
\end{equation} 

Using Sloan Digital Sky Survey (SDSS)  measurements of the galaxy power spectrum \cite{sdss1} and Wilkinson Microwave Anisotropy Probe (WMAP) \cite{spergelwmap1} 1$^{\rm st}$-year measurements of the CMB angular power spectrum, Refs.~\cite{hanraffelt,raffeltwong,hann5} derived limits of $m_{\rm a}\lsim 1~{\rm eV}$. We map these results into the 
$\left(\Omega_{a}h^{2},\lambda_{\rm fs}\right)$ plane.

We calculate $\Omega_{\rm a}\left(T_{\rm rh},m_{\rm a}\right)h^{2}$ and $\lambda_{\rm fs}\left(T_{\rm rh},m_{\rm a}\right)$ for axions in LTR, and thus obtain the upper limit to the axion mass as a function of $T_{\rm rh}$, shown in Fig.~\ref{limits}. For this data set, the smallest length scale for which the galaxy  correlation function can be reliably probed is $\lambda_{\rm min}\equiv 40~h^{-1}~{\rm Mpc}$ \cite{reh,hanraffelt,raffeltwong}. For $T_{\rm rh}\lsim35~{\rm MeV}$, $\lambda_{\rm fs}<40~h^{-1}~{\rm Mpc}$, and this axion mass constraint is lifted. At high $T_{\rm rh}$, the constraint from LSS/CMB data ($\Omega_{\rm a}h^{2}\lsim0.006$) supercedes the constraint $\Omega_{\rm a}h^2\lsim\Omega_{\rm m}h^{2}$.

Future instruments, such as the Large Synoptic Survey Telescope (LSST), will measure the matter power-spectrum with unprecedented precision ($\Delta P/P \sim 0.01)$ \cite{lsstzhan,lsstintro}. This order of magnitude improvement over past work \cite{sdss2,sdsslrg} leads to the improved sensitivity shown by the dotted line in Fig.~\ref{limits}. To estimate possible constraints to axions from LSST measurements of the power spectrum, we recalculated our limits using the approximate scaling $\Delta P/P\simeq -8\Omega_{\rm a}/\Omega_{\rm m}$, assuming $\Delta P/P \sim 10^{-2}$ for $\lambda>\lambda_{\rm min}=40~h^{-1}~{\rm Mpc}$.

We also estimate the possible improvement offered by including information on smaller scales ($\lambda_{\rm min}\sim 1~h^{-1}~{\rm Mpc}$), as probed by measurements of the Lyman-$\alpha$ flux power spectrum \cite{viel}, also shown in Fig.~\ref{limits}. This is indicated by the dashed line in Fig.~\ref{limits}.  We can see that higher $m_{\rm a}$ and lower $T_{\rm rh}$ values are probed because of information on smaller length scales.

\section{Axions, LTR, and BBN}
\begin{figure}[t]
\includegraphics[width=2.8in]{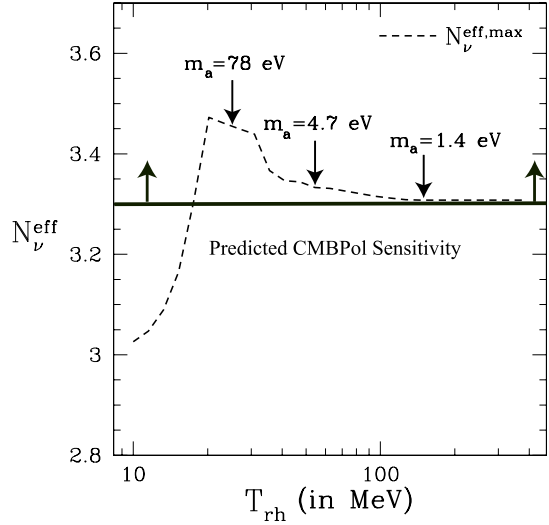}
\caption{Total effective neutrino number $N_{\nu}^{\rm eff,\rm max}$ for axions with masses saturating the tightest bound on axion masses from Fig.~\ref{limits}. The thick black line indicates the anticipated sensitivity of CMBPol \cite{kapling} to $N_{\nu}^{\rm eff}$, through measurements of $Y_{\rm p}$.}
\label{neff_dependence}
\end{figure}

Future limits to axions may follow from constraints to the total density in relativistic particles at $T\sim 1~{\rm MeV}$. This is parameterized by the axionic contribution to the total effective neutrino number $N_{\nu}^{\rm eff}$ \cite{reh,changchoi}:
\begin{eqnarray}
N_{\nu}^{\rm eff}=3+\frac{4}{7}\left(\frac{43}{4}\right)^{4/3} \Psi \left(T_{\rm F}/T_{\rm rh}\right),~
\Psi \left(y\right)\sim\left\{\begin{array}{ll}
\left[g_{*,\rm rh}y^{5}\left(\frac{g_{*,\rm F}}{g_{*,\rm rh}}\right)^{2}-1\right]^{-4/3}
&\mbox{if $y\gg1$,}\\
\left[g_{*,\rm F}-1\right]^{-4/3}&\mbox{if $y\ll 1$}.
\end{array}\right.\label{neffans}
\end{eqnarray} 
For sufficiently high masses, the axionic contribution saturates to $\delta N_{\nu}^{\rm eff}=4/7$ at high $T_{\rm rh}$ \cite{changchoi}. In Fig.~\ref{neff_dependence}, we show $N_{\nu}^{\rm eff,\rm max}\left(T_{\rm rh}\right)$, evaluated at whichever $m_{\rm a}$ which saturates the best cosmological bound for a given $T_{\rm rh}$. 

A comparison between the abundance of $^{4}$He ($Y_{\rm p}$) and the predicted abundance from BBN places constraints $N_{\rm \nu}^{\rm eff}$ at $T \sim 1$ MeV \cite{cardell}; thus constraints to $^{4}$He abundances are also constraints on $m_{\rm a}$ and $T_{\rm rh}$. Here we apply the scaling relation \cite{steigmanreview}:
\begin{equation}\Delta N^{\rm eff}_{\nu} = \frac{43}{7}\left\{\left(6.25\Delta Y_{p}+1\right)^{2}-1\right\}.\end{equation}

Direct measurements of $Y_{\rm p}$, including a determination of $\Omega_b$ from CMB observations, lead to the 68\% confidence level upper limit of $N_{\nu}^{\rm eff} \leq 3.85$ \cite{cyburt,ichikawa2,ichikawa4}.  From Fig.~\ref{neff_dependence}, we see that this bound cannot constrain $m_{\rm a}$ or $T_{\rm rh}$.  If future measurements reduce systematic errors, constraints to $T_{\rm rh}$ will be obtained for the lighter-mass axions.

Constraints to $m_{\rm a}$ and $T_{\rm rh}$ may follow from CMB measurements of $Y_p$.  $^4$He affects CMB anisotropies by changing the ionization history of the universe \cite{trotta}. CMBPol (a proposed future CMB polarization experiment) is expected to approach $\Delta Y_p = 0.0039$, leading to the sensitivity limit $N_{\nu}^{\rm eff} \leq 3.30$ \cite{trotta,kapling,ichikawa4,hut,cmbp}. As shown in Fig.~\ref{neff_dependence}, for $T_{\rm rh}\gsim 15~{\rm MeV}$, such measurements of $Y_{\rm p}$ may impose stringent limits on the axion mass. Also, if axions with mass in the $\rm eV$ range are directly detected, $Y_{p}$ might impose an \textit{upper} limit to $T_{\rm rh}$.

\section{Conclusions}
LTR suppresses the abundance of thermally-produced axions, once  $T_{\rm rh}\sim 50~{\rm MeV}$, as a result of dramatic entropy production. The cosmologically allowed window for $m_{\rm a}$ is extended as a result. Future probes of the matter power spectrum or the primordial helium abundance may definitively explore some of this parameter space.

%
%%%%%%%%%%%%%%%%%%%%%%%%%%%%%%%%%%%%%%%%%%%%%%%%%
%%% BACKMATTER
%%%%%%%%%%%%%%%%%%%%%%%%%%%%%%%%%%%%%%%%%%%%%%%%%

\begin{theacknowledgments}
D.G. was supported by the Gordon and Betty Moore Foundation and thanks the organizers of DM08. T.L.S. and M.K. were supported by DoE DE-FG03-92-ER40701 and the Gordon and Betty Moore Foundation. 
\end{theacknowledgments}

%%%%%%%%%%%%%%%%%%%%%%%%%%%%%%%%%%%%%%%%%%%%%%%%%
%%% The bibliography can be prepared using the BibTeX program or
%%% manually.
%%%
%%% The code below assumes that BibTeX is used.  If the bibliography is
%%% produced without BibTeX comment out the following lines and see the
%%% aipguide.pdf for further information.
%%%
%%% For your convenience a manually coded example is appended
%%% after the \end{document}
%%%%%%%%%%%%%%%%%%%%%%%%%%%%%%%%%%%%%%%%%%%%%%%%%

%%%%%%%%%%%%%%%%%%%%%%%%%%%%%%%%%%%%%%%%%%%%%%%%%
%%% You may have to change the BibTeX style below, depending on your
%%% setup or preferences.
%%%
%%%
%%% For The AIP proceedings layouts use either
%%%%%%%%%%%%%%%%%%%%%%%%%%%%%%%%%%%%%%%%%%%%%

%\bibliographystyle{aipproc}   % if natbib is available
%\bibliography{dm08_grin_proc.bib}

\end{document}